\documentclass[a4paper,
               biblatex,     
               keeplastbox,   
               hyphens,      
               ]{jacow}
\usepackage{pdfpages,multirow,ragged2e} %
\makeatletter%
	\ifboolexpr{bool{xetex}}
	 {\renewcommand{\Gin@extensions}{.pdf,%
	                    .png,.jpg,.bmp,.pict,.tif,.psd,.mac,.sga,.tga,.gif,%
	                    .eps,.ps,%
	                    }}{}
\makeatother

\ifboolexpr{bool{xetex} or bool{luatex}} 
 {}                                      

\usepackage[USenglish]{babel}

\usepackage{caption, subcaption}
\usepackage{lipsum}
\usepackage{hyperref}
\usepackage{IEEEtrantools}
\usepackage{tabularx}
\usepackage{tabulary}

\ifboolexpr{bool{jacowbiblatex}}%
 {%
  \addbibresource{references.bib}
 }{}
\begin{document}

\title{Update on the LLRF operations status at the European XFEL}

\author{M. Diomede\thanks{marco.diomede@desy.de}, V. Ayvazyan, J. Branlard, M. Grecki, M. Hierholzer, M. Hoffmann, 
\\ B. Lautenschlager, S. Pfeiffer, C. Schmidt, N. Walker,
\\ Deutsches Elektronen-Synchrotron DESY, Notkestr. 85, 22607 Hamburg, Germany}
	
\maketitle

\begin{abstract}
   The European XFEL (EuXFEL) is a free-electron laser in the X-ray range for users. Its high availability is one of the key aspects of the machine. In 2022, it entered in the sixth year of operation. The EuXFEL linac is based on the TESLA superconducting RF technology, operating at 1.3 GHz with an RF pulse length of \SI{650}{\micro s} and a repetition rate of 10 Hz. The LLRF system is based on the MicroTCA standard and relies on a high level of automation. In this contribution, we review the LLRF operation at the EuXFEL and the development of new tools and features to improve the monitoring and extend the usability of the LLRF system.
\end{abstract}

\section{THE EUROPEAN XFEL}
The European XFEL is a hard X-ray free-electron laser (FEL) based on a high-electron-energy superconducting linear accelerator \cite{Decking2020AAccelerator}. With different linac energies ranging from 8 to 17.5 GeV and variable-gap undulators, photon energies from 0.25 to 25 keV can be covered. 

The baseline frequency of the RF system is 1.3 GHz. The linac is made up of a normal conducting gun followed by an accelerating module (named A1) and a third harmonic lineariser module (called AH1) operating at 3.9 GHz. The linac is then completed by other 24 RF stations (A2-A25), each of them fed by one 10 MW multibeam klystron \cite{Bousonville2021EuropeanOperation}. The regular RF station is made up of 4 cryomodules (with the exception of A1 and AH1 that have only one accelerating module each), where every module is made up of 8 standing wave superconducting TESLA cavities \cite{Aune2000SuperconductingCavities}.

The multicavity LLRF system is based on the MicroTCA.4 standard (MTCA). It provides RMS field stability
better than 0.01\% in amplitude and 0.01 deg in phase at the 1.3 GHz cavity operating frequency \cite{Branlard2012TheSystem}. For every regular RF station, it is split into manager and subordinate crates as each of them can process the signals from up to 2 cryomodules. For this reason, A1 and AH1 can share the same MTCA crate.

\section{LLRF STATISTICS}
At the EuXFEL, there are currently 4 reference energy settings: 8, 11.5, 14 and 16.3 GeV. The first three are comprised in the so-called "reduced-energy configuration (RE)", while the latter in the "high-energy configuration (HE)". The differences between the two configurations are the modulator high-voltage of some RF stations and the average gradient in the cavities. Switching between the three beam energies of the reduced-energy configuration is done by changing the off-crest RF phase by means of the Linac Energy Manager server \cite{Hensler2017RF-EnergyXFEL}.

Running at high-energy implies a higher accelerating gradient in the cavities, a higher dark current radiation and then a higher incidence of Single Event Upsets (SEUs) that can affect the MTCA boards, causing station trips. 

In order to maintain the full functionality of the LLRF system, a weekly maintenance is performed every Monday for 1 hour. Due to the complexity and extension of the LLRF system at the EuXFEL, having it always in nominal running condition (i.e. every single aspect of it at the desired status or setting) might be not easy. Most important properties of the LLRF system that are regularly checked in order to guarantee full functionality comprise: status of the cavity tuners and RF power coupler motors, resonance frequency of the cavities and loaded quality factor ($Q_l$), status of the tuning piezos, reference phase, status of the MTCA crates, LLRF feedbacks, cavity power limiters, quench protection, drift compensation modules, control system servers, power supplies of the racks, etc... Thus, typical operations during weekly maintenance include:
\begin{itemize}
    \item Power-cycle of faulty MTCA boards, CPUs, crates;
    \item Piezo relaxation (reduction of piezo DC voltage to extend their lifetime);
    \item $Q_l$ adjustment;
    \item Beam based calibration.
\end{itemize}

\begin{figure}[!htb]
   \centering
   \includegraphics*[width=\columnwidth]{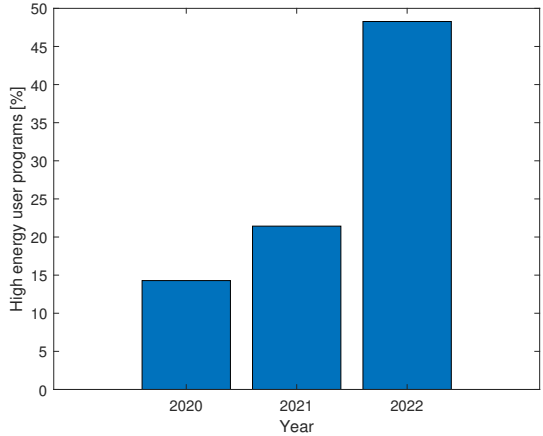}
   \caption{High-energy user programs through the years.}
   \label{fig:HE_stats}
\end{figure}

The 16.3 GeV energy is highly requested by the FEL users. Indeed, in 2022, EuXFEL will run at high-energy for almost 50\% of the user time (Fig. \ref{fig:HE_stats}). 

\begin{figure}[!htb]
   \centering
   \includegraphics*[width=\columnwidth]{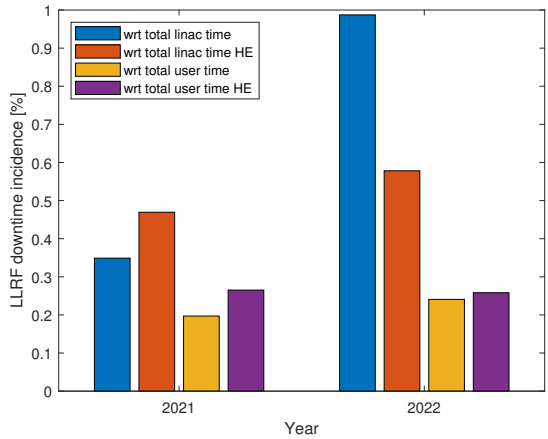}
   \caption{LLRF downtime of the past two years (until 18/09/2022).}
   \label{fig:LLRF_stats}
\end{figure}

In Table \ref{tab:LLRF_stats} and Fig. \ref{fig:LLRF_stats}, the LLRF downtime incidence with respect to the total linac time and total user time is shown. In addition to the user time, the total linac time is given by setup, tuning and facility development time. From the statistics, it is possible to observe that running in high-energy increases the contribution of the LLRF system to the total linac downtime. Nevertheless, during user time, this contribution is limited to a value that is lower than 0.3\%. A very good result, considering the total length of the linac and the radiation prone environment (the racks are located underneath the cryomodules in the same tunnel). During reduced-energy runs, it is possible to perform tests on a few RF stations placing them "off-beam", i.e. shifting their timing with respect to beam. These studies allow to discover field emitters and quenching cavities. The solution applied in these cases is to reduce the maximum voltage of the affected RF station or to detune the affected cavity.

\begin{table}[!htb]
    \caption{LLRF downtime of the past two years (until 18/09/2022).}
    \label{tab:LLRF_stats}	
	\begin{tabulary}{\columnwidth}{LCCCCR}
		\hline
		Year& Total linac time [days]& Total LLRF downtime [hours]& Total user time [days]& User LLRF downtime [hours]& User HE weeks\\
		\hline
        2021& 277.36& 23.23& 168.00& 7.94& 6\\
        2022& 236.21& 55.97& 136.00& 7.86& 14\\
        \hline
		\end{tabulary}
\end{table}

\section{NEW TOOLS AND FEATURES}
Due to the large scale of the accelerator, a high degree of automation is provided to assist operators and system experts, including exception detection and handling. In the past months, several new tools and features have been introduced at the EuXFEL to further improve the monitoring and extend the usability of the LLRF system. 

\subsection{LLRF status tool}
As mentioned in the previous section, maintenance of the LLRF system is regularly performed. Having an instantaneous picture of such a complex and distributed system is very helpful. Including in a single jddd (Java Data Display program for the Distributed Object Oriented Control System) \cite{Sombrowski2007JDDD:XFEL} panel information about the status of MTCA crates, LLRF controllers and servers, cavity signals, external hardware units, coupler and cavity tuners is almost impossible. For this reason, a MATLAB script has been developed to check hundreds of Distributed Object Oriented Control System (DOOCS) (\cite{Grygiel1996DOOCS:Workstations}) properties in about 30 seconds. By means of the GUI, it is also possible to check a reduced number of properties grouped by LLRF subsystem and obtain a quicker output. The tool generates a textual report and the run of the script is daily scheduled through cron jobs with automatic emails sent to the experts. This tool allows to have an updated status of the system so that everything can be fixed during maintenance time.



\subsection{Cavity $Q_l$ tuner tool}
\begin{figure}[!htb]
   \centering
   \includegraphics*[width=\columnwidth]{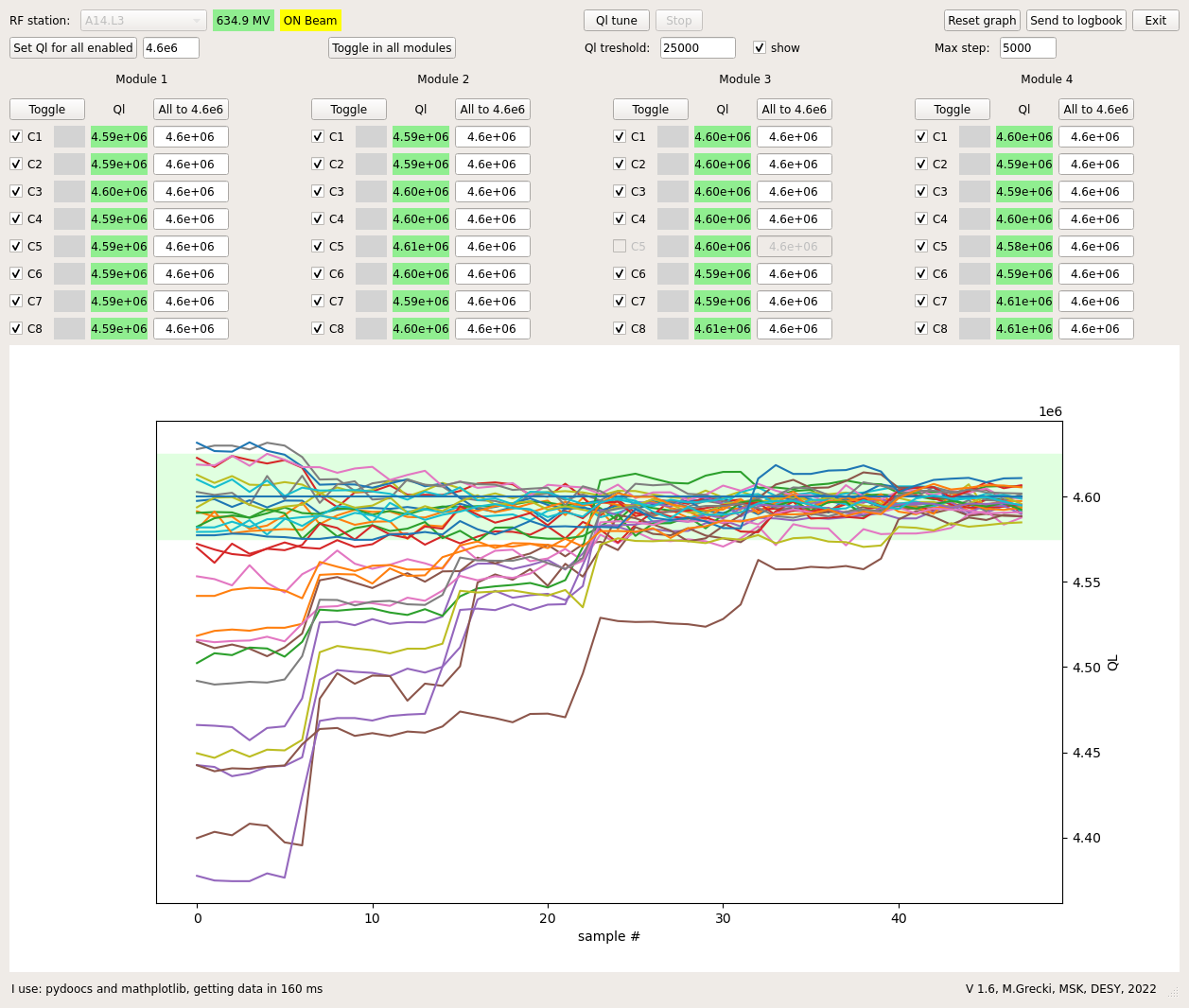}
   \caption{Cavity $Q_l$ tuner tool.}
   \label{fig:cavity_Ql_tuner}
\end{figure}

A new cavity $Q_l$ tuner Python script has been recently developed and deployed at the EuXFEL. As already mentioned in the previous section, the $Q_l$ adjustment is part of the activities performed during LLRF maintenance. In particular, it is performed at every RF station when switched from one energy configuration to another. The approach of the script is similar to the already existing cavity resonance frequency tuner script. With respect to the previously adopted script \cite{Cichalewski2012SuperconductingFLASH}, this new version directly acts on all the modules of an RF station, reducing the required tuning time. A picture of the cavity $Q_l$ tuner's GUI is shown in Fig. \ref{fig:cavity_Ql_tuner}. Through its GUI, it is possible to select or exclude single cavities and modules. It is also possible to redefine the global goal value of the station or set individual cavity values. Finally, it is possible to set the tolerance threshold and the maximum number of tuner steps the algorithm can change at each iteration.

\subsection{MTCA hot-plug feature}
Running at high-energy increases the rate of SEUs. SEUs lead to an RF station trip if probe boards are affected. RF stations usually survive the whole user week run before maintenance if forward and reflected boards are hit. This is because the probe signals are used for field regulation (vector sum of the RF station), while forward and reflected signals are used for diagnostic purposes to calculate the cavity detuning and the $Q_l$, and to compute the virtual probe that can be used in case of failure of the probe signal \cite{Pfeiffer2015VirtualSignals}. In the past, the way to fix faulty boards was to power-cycle the MTCA crate. Recently, the hot-plug support has been enabled in all MicroTCA Carrier Hubs (MCHs). This now allows to just power-cycle the faulty board and rescan the PCIe bus on the relative CPU. The whole procedure can be performed directly from the dedicated jddd panel. The procedure is much faster than the past, as it requires around 1 minute in total while the power-cycle of the crate requires 5 minutes before all the servers are up and running due to the CPU restart. In addition, a crate power-cycle often results in jumps of the reference phase and thus the re-calibration of the cavity signals are required. We are also experiencing wrong initialization of the down converter board attenuators for which re-initialization is sometimes necessary. So, power-cycling only the faulty board reduces the necessary checks on the RF station during recovery.

\subsection{Beam regions}
\begin{figure*}[!hbt]
   \centering
   \includegraphics*[width=0.7\textwidth]{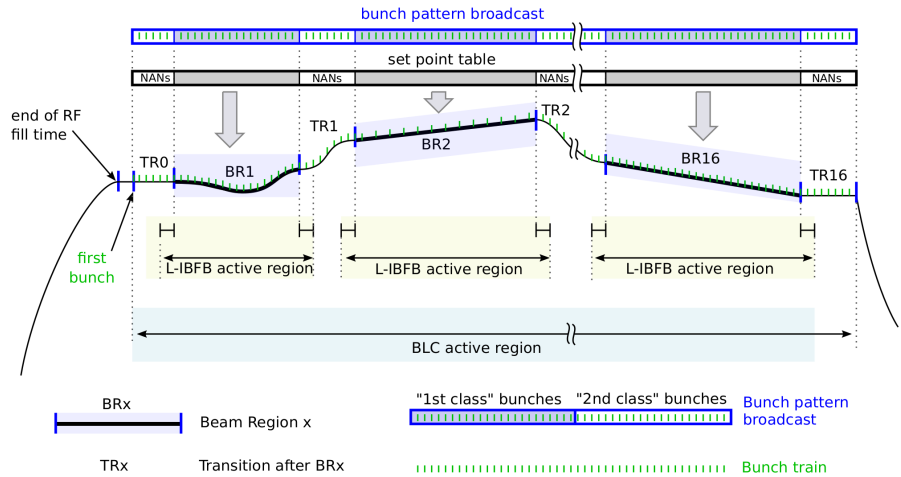}
   \caption{Schematic view of the beam regions.}
   \label{fig:beam_regions}
\end{figure*}

During the summer 2022 shutdown, the so-called "beam regions" (BRs) feature has been released at EuXFEL. It represents a change with respect to the previous multi flat tops scheme \cite{Ayvazyan2008AlternatingFLASH}. The maximum number of flat tops was limited to 3, while now there can be up to 16 beam regions. The beam regions can now have arbitrary shape and are not just limited to a constant amplitude. The beam regions are separated by transition regions (TRs). A schematic view of the concept is shown in Fig. \ref{fig:beam_regions}. The beam is present over the entire RF flat top but only the bunches accelerated during the BR will be used to lase ("first class" bunches). For this reason, the longitudinal intra bunch train feedback (L-IBFB) stabilizing the bunch arrival time and compression is active during the BRs but switched off during the TRs. The beam loading compensation (BLC) is active throughout. The shape of the BRs is provided by a high-level server, while the set point transitions between BRs are calculated by the LLRF controller server.

When the beam regions are enabled, the RF stations run in the so-called "table mode", while, when disabled, in the "scalar mode", which corresponds to just one flat top. The set point tables are generated by the llrfCtrl server. The llrfCtrl server rejects beam region requests that correspond to exceeding the vector limit of \SI{0.7}{MV/\micro s}. This is considered to be the maximum deviation, in the complex plane, that is sustainable by the klystron bandwidth. It is possible to easily switch between the table mode and scalar mode by means of the main LLRF panel. The number of beam regions is determined by the Bunch Pattern Builder server \cite{Frohlich2019Multi-beamlineXFEL}. The RF station is recovered in scalar mode after an RF trip. At the moment, at the EuXFEL, the stations from A1 to A5 (i.e. Injector 1, Linac 1 and Linac 2) are running in table mode with up to two beam regions with simple flat top tables. Stations from A6 to A25 (i.e. Linac 3) are operated in scalar mode. Results of flexible operating modes at the EuXFEL, also using the beam regions, are reported in \cite{Guetg2022FlexibleEuXFEL}.


\section{MARWIN XTL RADIATION SCANS}
\begin{figure*}[!htb]
   \centering
   \includegraphics*[width=0.9\textwidth]{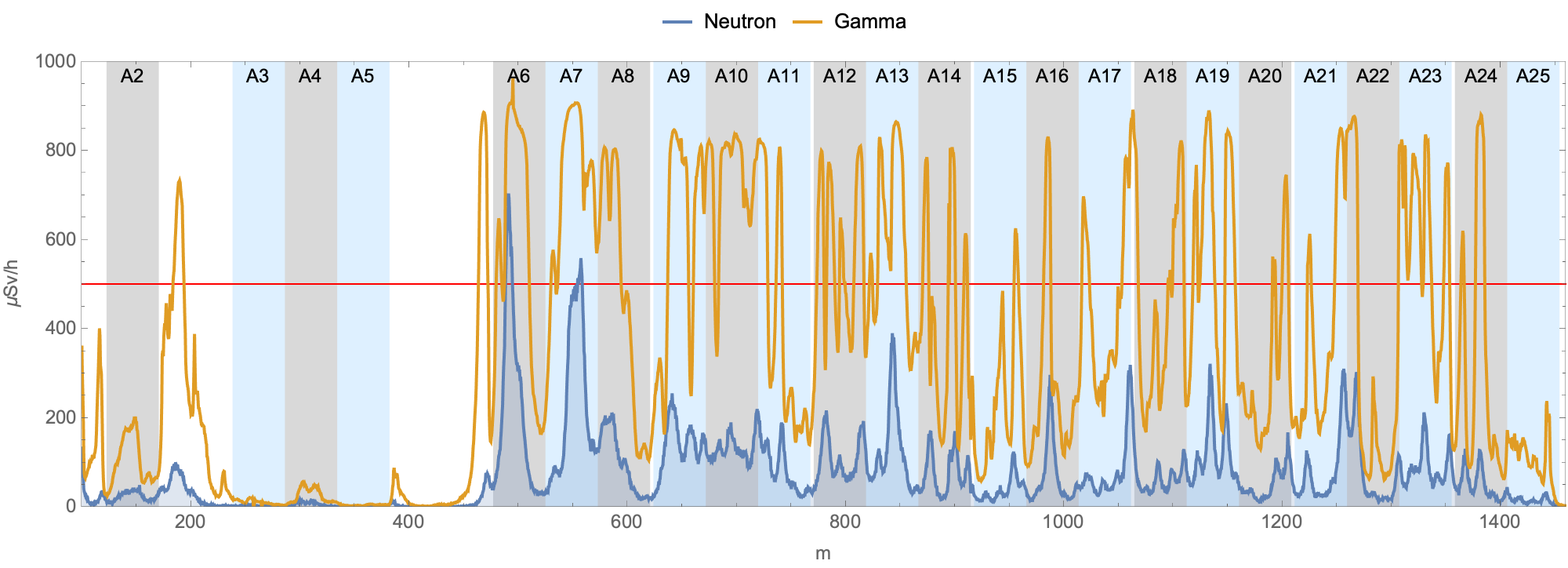}
   \caption{Typical radiation profile of the XTL tunnel during high-energy run.}
   \label{fig:XTL_scan}
\end{figure*}

During high-energy runs, weekly radiation scans of the XTL tunnel (i.e. RF stations from A2 to A25) are performed. The scans are run by means of the MARWIN moving robot \cite{Dehne2018Marwin:Environment}. The robot is capable of autonomous motion inside the tunnel and measures instantaneous gamma and neutron radiation levels. If the neutron radiation level of a station repetitively exceeds the administrative limit of \SI{500}{\micro Sv/h}, its accelerating gradient is reduced or the emitting cavity identified and detuned. The typical radiation profile of the XTL tunnel during high-energy run is reported in Fig. \ref{fig:XTL_scan}. In the plot, it is possible to observe a high radiation level at the RF station A6. The station has always been affected by cavity field emitters producing dark current. However, in this particular case, the radiation was mainly due to a non optimal setup of the bunch compressor BC2 that lead to beam losses at the chicane. Over the years, several other field emitters have been discovered \cite{Branlard2021FourXFEL}.



\section{SUMMARY}
An overview of the LLRF operations at the European XFEL is given in this paper. The statistics show that the downtime incidence of the LLRF system is maintained at a level <0.3\% with respect to the user linac time even though, in 2022, the high-energy user programs have risen to almost 50\% of the user time. Due to the large scale of the accelerator, a high degree of automation is provided to assist operators and system experts, including exception detection and handling. New tools and features have been developed to further improve the monitoring and extend the usability of the LLRF system. Among them, there is the LLRF status tool that checks hundreds of DOOCS properties and provides the health status of the whole LLRF system with its subsystems. Another recent new tool is the cavity $Q_l$ tuner script. This script, together with its GUI, allows to tune the $Q_l$ of all the cavities of an RF stations in a few tens of seconds. Enabling the MTCA hot-plug feature allowed to reduce the time to recover an RF station in case of a faulty board due to SEU. The beam regions feature introduced the possibility to have arbitrary RF shapes in amplitude and phase. Radiation scans are performed weekly during high-energy runs by means of the moving robot MARWIN in order to keep the neutron radiation level of the linac below \SI{500}{\micro Sv/h}.

%
%
\ifboolexpr{bool{jacowbiblatex}}%
{\printbibliography}%

@article{Decking2020AAccelerator,
    title = {{A MHz-repetition-rate hard X-ray free-electron laser driven by a superconducting linear accelerator}},
    year = {2020},
    journal = {Nature Photonics},
    author = {Decking, W. and Abeghyan, S. and Abramian, P. and Abramsky, A. and Aguirre, A. and Albrecht, C. and Alou, P. and Altarelli, M. and Altmann, P. and Amyan, K. and Anashin, V. and Apostolov, E. and Appel, K. and Auguste, D. and Ayvazyan, V. and Baark, S. and Babies, F. and Baboi, N. and Bak, P. and Balandin, V. and Baldinger, R. and Baranasic, B. and Barbanotti, S. and Belikov, O. and Belokurov, V. and Belova, L. and Belyakov, V. and Berry, S. and Bertucci, M. and Beutner, B. and Block, A. and Bl{\"{o}}cher, M. and B{\"{o}}ckmann, T. and Bohm, C. and B{\"{o}}hnert, M. and Bondar, V. and Bondarchuk, E. and Bonezzi, M. and Borowiec, P. and B{\"{o}}sch, C. and B{\"{o}}senberg, U. and Bosotti, A. and B{\"{o}}spflug, R. and Bousonville, M. and Boyd, E. and Bozhko, Y. and Brand, A. and Branlard, J. and Briechle, S. and Brinker, F. and Brinker, S. and Brinkmann, R. and Brockhauser, S. and Brovko, O. and Br{\"{u}}ck, H. and Br{\"{u}}dgam, A. and Butkowski, L. and B{\"{u}}ttner, T. and Calero, J. and Castro-Carballo, E. and Cattalanotto, G. and Charrier, J. and Chen, J. and Cherepenko, A. and Cheskidov, V. and Chiodini, M. and Chong, A. and Choroba, S. and Chorowski, M. and Churanov, D. and Cichalewski, W. and Clausen, M. and Clement, W. and Clou{\'{e}}, C. and Cobos, J. A. and Coppola, N. and Cunis, S. and Czuba, K. and Czwalinna, M. and D’Almagne, B. and Dammann, J. and Danared, H. and de Zubiaurre Wagner, A. and Delfs, A. and Delfs, T. and Dietrich, F. and Dietrich, T. and Dohlus, M. and Dommach, M. and Donat, A. and Dong, X. and Doynikov, N. and Dressel, M. and Duda, M. and Duda, P. and Eckoldt, H. and Ehsan, W. and Eidam, J. and Eints, F. and Engling, C. and Englisch, U. and Ermakov, A. and Escherich, K. and Eschke, J. and Saldin, E. and Faesing, M. and Fallou, A. and Felber, M. and Fenner, M. and Fernandes, B. and Fern{\'{a}}ndez, J. M. and Feuker, S. and Filippakopoulos, K. and Floettmann, K. and Fogel, V. and Fontaine, M. and Franc{\'{e}}s, A. and Martin, I. Freijo and Freund, W. and Freyermuth, T. and Friedland, M. and Fr{\"{o}}hlich, L. and Fusetti, M. and Fydrych, J. and Gallas, A. and Garc{\'{i}}a, O. and Garcia-Tabares, L. and Geloni, G. and Gerasimova, N. and Gerth, C. and Ge{\ss}ler, P. and Gharibyan, V. and Gloor, M. and G{\l}owinkowski, J. and Goessel, A. and Go{\l}{\c{e}}biewski, Z. and Golubeva, N. and Grabowski, W. and Graeff, W. and Grebentsov, A. and Grecki, M. and Grevsmuehl, T. and Gross, M. and Grosse-Wortmann, U. and Gr{\"{u}}nert, J. and Grunewald, S. and Grzegory, P. and Feng, G. and Guler, H. and Gusev, G. and Gutierrez, J. L. and Hagge, L. and Hamberg, M. and Hanneken, R. and Harms, E. and Hartl, I. and Hauberg, A. and Hauf, S. and Hauschildt, J. and Hauser, J. and Havlicek, J. and Hedqvist, A. and Heidbrook, N. and Hellberg, F. and Henning, D. and Hensler, O. and Hermann, T. and Hidv{\'{e}}gi, A. and Hierholzer, M. and Hintz, H. and Hoffmann, F. and Hoffmann, Markus and Hoffmann, Matthias and Holler, Y. and H{\"{u}}ning, M. and Ignatenko, A. and Ilchen, M. and Iluk, A. and Iversen, J. and Iversen, J. and Izquierdo, M. and Jachmann, L. and Jardon, N. and Jastrow, U. and Jensch, K. and Jensen, J. and Je{\.{z}}abek, M. and Jidda, M. and Jin, H. and Johannson, N. and Jonas, R. and Kaabi, W. and Kaefer, D. and Kammering, R. and Kapitza, H. and Karabekyan, S. and Karstensen, S. and Kasprzak, K. and Katalev, V. and Keese, D. and Keil, B. and Kholopov, M. and Killenberger, M. and Kitaev, B. and Klimchenko, Y. and Klos, R. and Knebel, L. and Koch, A. and Koepke, M. and K{\"{o}}hler, S. and K{\"{o}}hler, W. and Kohlstrunk, N. and Konopkova, Z. and Konstantinov, A. and Kook, W. and Koprek, W. and K{\"{o}}rfer, M. and Korth, O. and Kosarev, A. and Kosi{\'{n}}ski, K. and Kostin, D. and Kot, Y. and Kotarba, A. and Kozak, T. and Kozak, V. and Kramert, R. and Krasilnikov, M. and Krasnov, A. and Krause, B. and Kravchuk, L. and Krebs, O. and Kretschmer, R. and Kreutzkamp, J. and Kr{\"{o}}plin, O. and Krzysik, K. and Kube, G. and Kuehn, H. and Kujala, N. and Kulikov, V. and Kuzminych, V. and La Civita, D. and Lacroix, M. and Lamb, T. and Lancetov, A. and Larsson, M. and Le Pinvidic, D. and Lederer, S. and Lensch, T. and Lenz, D. and Leuschner, A. and Levenhagen, F. and Li, Y. and Liebing, J. and Lilje, L. and Limberg, T. and Lipka, D. and List, B. and Liu, J. and Liu, S. and Lorbeer, B. and Lorkiewicz, J. and Lu, H. H. and Ludwig, F. and Machau, K. and Maciocha, W. and Madec, C. and Magueur, C. and Maiano, C. and Maksimova, I. and Malcher, K. and Maltezopoulos, T. and Mamoshkina, E. and Manschwetus, B. and Marcellini, F. and Marinkovic, G. and Martinez, T. and Martirosyan, H. and Maschmann, W. and Maslov, M. and Matheisen, A. and Mavric, U. and Mei{\ss}ner, J. and Meissner, K. and Messerschmidt, M. and Meyners, N. and Michalski, G. and Michelato, P. and Mildner, N. and Moe, M. and Moglia, F. and Mohr, C. and Mohr, S. and M{\"{o}}ller, W. and Mommerz, M. and Monaco, L. and Montiel, C. and Moretti, M. and Morozov, I. and Morozov, P. and Mross, D. and Mueller, J. and M{\"{u}}ller, C. and M{\"{u}}ller, J. and M{\"{u}}ller, K. and Munilla, J. and M{\"{u}}nnich, A. and Muratov, V. and Napoly, O. and N{\"{a}}ser, B. and Nefedov, N. and Neumann, Reinhard and Neumann, Rudolf and Ngada, N. and Noelle, D. and Obier, F. and Okunev, I. and Oliver, J. A. and Omet, M. and Oppelt, A. and Ottmar, A. and Oublaid, M. and Pagani, C. and Paparella, R. and Paramonov, V. and Peitzmann, C. and Penning, J. and Perus, A. and Peters, F. and Petersen, B. and Petrov, A. and Petrov, I. and Pfeiffer, S. and Pfl{\"{u}}ger, J. and Philipp, S. and Pienaud, Y. and Pierini, P. and Pivovarov, S. and Planas, M. and P{\l}awski, E. and Pohl, M. and Polinski, J. and Popov, V. and Prat, S. and Prenting, J. and Priebe, G. and Pryschelski, H. and Przygoda, K. and Pyata, E. and Racky, B. and Rathjen, A. and Ratuschni, W. and Regnaud-Campderros, S. and Rehlich, K. and Reschke, D. and Robson, C. and Roever, J. and Roggli, M. and Rothenburg, J. and Rusi{\'{n}}ski, E. and Rybaniec, R. and Sahling, H. and Salmani, M. and Samoylova, L. and Sanzone, D. and Saretzki, F. and Sawlanski, O. and Schaffran, J. and Schlarb, H. and Schl{\"{o}}sser, M. and Schlott, V. and Schmidt, C. and Schmidt-Foehre, F. and Schmitz, M. and Schm{\"{o}}kel, M. and Schnautz, T. and Schneidmiller, E. and Scholz, M. and Sch{\"{o}}neburg, B. and Schultze, J. and Schulz, C. and Schwarz, A. and Sekutowicz, J. and Sellmann, D. and Semenov, E. and Serkez, S. and Sertore, D. and Shehzad, N. and Shemarykin, P. and Shi, L. and Sienkiewicz, M. and Sikora, D. and Sikorski, M. and Silenzi, A. and Simon, C. and Singer, W. and Singer, X. and Sinn, H. and Sinram, K. and Skvorodnev, N. and Smirnow, P. and Sommer, T. and Sorokin, A. and Stadler, M. and Steckel, M. and Steffen, B. and Steinhau-K{\"{u}}hl, N. and Stephan, F. and Stodulski, M. and Stolper, M. and Sulimov, A. and Susen, R. and {\'{S}}wierblewski, J. and Sydlo, C. and Syresin, E. and Sytchev, V. and Szuba, J. and Tesch, N. and Thie, J. and Thiebault, A. and Tiedtke, K. and Tischhauser, D. and Tolkiehn, J. and Tomin, S. and Tonisch, F. and Toral, F. and Torbin, I. and Trapp, A. and Treyer, D. and Trowitzsch, G. and Trublet, T. and Tschentscher, T. and Ullrich, F. and Vannoni, M. and Varela, P. and Varghese, G. and Vashchenko, G. and Vasic, M. and Vazquez-Velez, C. and Verguet, A. and Vilcins-Czvitkovits, S. and Villanueva, R. and Visentin, B. and Viti, M. and Vogel, E. and Volobuev, E. and Wagner, R. and Walker, N. and Wamsat, T. and Weddig, H. and Weichert, G. and Weise, H. and Wenndorf, R. and Werner, M. and Wichmann, R. and Wiebers, C. and Wiencek, M. and Wilksen, T. and Will, I. and Winkelmann, L. and Winkowski, M. and Wittenburg, K. and Witzig, A. and Wlk, P. and Wohlenberg, T. and Wojciechowski, M. and Wolff-Fabris, F. and Wrochna, G. and Wrona, K. and Yakopov, M. and Yang, B. and Yang, F. and Yurkov, M. and Zagorodnov, I. and Zalden, P. and Zavadtsev, A. and Zavadtsev, D. and Zhirnov, A. and Zhukov, A. and Ziemann, V. and Zolotov, A. and Zolotukhina, N. and Zummack, F. and Zybin, D.},
    number = {6},
    volume = {14},
    doi = {10.1038/s41566-020-0607-z}
}

@inproceedings{Grygiel1996DOOCS:Workstations,
    title = {{DOOCS: a Distributed Object Oriented Control system on PC’s and workstations}},
    year = {1996},
    booktitle = {PCaPAC, Hamburg, Germany},
    author = {Grygiel, G and Hensler, O and Rehlich, K}
}

@inproceedings{Bousonville2021EuropeanOperation,
    title = {{European XFEL High-Power RF System - the First 4 Years of Operation}},
    year = {2021},
    booktitle = {IPAC, Campinas, Brazil},
    author = {Bousonville, M and Choroba, S and Grevsm{\"{u}}hl, T and G{\"{o}}ller, S and Hauberg, A and Katalev, V V and Machau, K and Vogel, V and Yildirim, B},
    doi = {10.18429/JACoW-IPAC2021-WEPAB045}
}

@inproceedings{Branlard2021FourXFEL,
    title = {{Four Years of Successful Operation of the European XFEL}},
    year = {2021},
    booktitle = {SRF, East Lansing, MI, USA},
    author = {Branlard, J and Choroba, S and Grecki, M K and Kostin, D and K{\"{o}}pke, S and N{\"{o}}lle, D and Vogel, V and Walker, N and Wiesenberg, S},
    doi = {10.18429/JACoW-SRF2021-MOOFAV06}
}

@article{Sombrowski2007JDDD:XFEL,
    title = {{JDDD: a Java DOOCS data display for the XFEL}},
    year = {2007},
    journal = {ICALEPCS, Knoxville, TN, USA},
    author = {Sombrowski, Elke and Petrosyan, A and Rehlich, K and Tege, P}
}

@article{Dehne2018Marwin:Environment,
    title = {{Marwin: Localization of an inspection robot in a radiation-exposed environment}},
    year = {2018},
    journal = {Advances in Science, Technology and Engineering Systems},
    author = {Dehne, Andre and M{\"{o}}ller, Nantwin and Hermes, Thorsten},
    number = {4},
    volume = {3},
    doi = {10.25046/aj030436}
}

@inproceedings{Frohlich2019Multi-beamlineXFEL,
    title = {{Multi-beamline operation at the European XFEL}},
    year = {2019},
    booktitle = {FEL, Hamburg, Germany},
    author = {Fr{\"{o}}hlich, L. and Aghababyan, A. and Balandin, V. and Beutner, B. and Brinker, F. and Decking, W. and Golubeva, N. and Hensler, O. and Janik, Y. and Kammering, R. and Kay, H. and Limberg, T. and Liu, S. and N{\"{o}}lle, D. and Obier, F. and Omet, M. and Scholz, M. and Wamsat, T. and Wilksen, T. and Wortmann, J.},
    doi = {10.18429/JACoW-FEL2019-WEP008}
}

@article{Hensler2017RF-EnergyXFEL,
    title = {{RF-Energy Management for the European XFEL}},
    year = {2017},
    journal = {ICALEPCS, Barcelona, Spain},
    author = {Hensler, O and {others}},
    doi = {10.18429/JACoW-ICALEPCS2017-THMPA04}
}

@inproceedings{Cichalewski2012SuperconductingFLASH,
    title = {{Superconducting cavities automatic loaded quality factor control at FLASH}},
    year = {2012},
    booktitle = {2012 18th IEEE-NPSS Real Time Conference},
    author = {Cichalewski, W. and Branlard, J. and Schlarb, H. and Carwardine, J. and Napieralski, A.},
    doi = {10.1109/RTC.2012.6418094}
}

@article{Aune2000SuperconductingCavities,
    title = {{Superconducting TESLA cavities}},
    year = {2000},
    journal = {Physical Review Special Topics - Accelerators and Beams},
    author = {Aune, B. and Bandelmann, R. and Bloess, D. and Bonin, B. and Bosotti, A. and Champion, M. and Crawford, C. and Deppe, G. and Dwersteg, B. and Edwards, D. A. and Edwards, H. T. and Ferrario, M. and Fouaidy, M. and Gall, P. D. and Gamp, A. and G{\"{o}}ssel, A. and Graber, J. and Hubert, D. and H{\"{u}}ning, M. and Juillard, M. and Junquera, T. and Kaiser, H. and Kreps, G. and Kuchnir, M. and Lange, R. and Leenen, M. and Liepe, M. and Lilje, L. and Matheisen, A. and M{\"{o}}ller, W. D. and Mosnier, A. and Padamsee, H. and Pagani, C. and Pekeler, M. and Peters, H. B. and Peters, O. and Proch, D. and Rehlich, K. and Reschke, D. and Safa, H. and Schilcher, T. and Schm{\"{u}}ser, P. and Sekutowicz, J. and Simrock, S. and Singer, W. and Tigner, M. and Trines, D. and Twarowski, K. and Weichert, G. and Weisend, J. and Wojtkiewicz, J. and Wolff, S. and Zapfe, K.},
    number = {9},
    volume = {3},
    doi = {10.1103/physrevstab.3.092001}
}

@inproceedings{Branlard2012TheSystem,
    title = {{The European XFEL LLRF System}},
    year = {2012},
    booktitle = {IPAC, New Orleans, LA, USA},
    author = {Branlard, J and {others}},
    pages = {},
    publisher = {},
    language = {english}
}
{%

} 
\end{document}